\documentclass[twocolumn,showpacs,aps,prl,groupedaddress,superscriptaddress]{revtex4-1}

\usepackage{mathrsfs}
\usepackage{amssymb, amsbsy, amsmath, latexsym, dsfont, array, layout, mathrsfs, color}

\usepackage{mathtools}

\usepackage{amsmath,amssymb}
\usepackage{bm}
\usepackage{graphicx}
\usepackage{braket,dsfont}
\usepackage{color}
\usepackage[10pt]{moresize}
\usepackage{mathrsfs}
\usepackage{multirow}
\usepackage{hyperref}
\usepackage{changes}
\usepackage{mathdots}
\let\oldcaption\caption
\renewcommand{\caption}{\sffamily \oldcaption}

\usepackage{amsmath,amsfonts,amssymb}
\usepackage{wrapfig}
\usepackage{graphicx}
\usepackage{bbm}
\usepackage[normalem]{ulem}
\usepackage{color}

\begin{document}

\title{Nonclassical Preparation of Quantum Remote States}\date{\today}

\author{Shih-Hsuan Chen}
\affiliation{Department of Engineering Science, National Cheng Kung University, Tainan 70101, Taiwan}
\affiliation{Center for Quantum Frontiers of Research $\&$ Technology, National Cheng Kung University, Tainan 70101, Taiwan}

\author{Yu-Chien Kao}
\affiliation{Department of Engineering Science, National Cheng Kung University, Tainan 70101, Taiwan}
\affiliation{Center for Quantum Frontiers of Research $\&$ Technology, National Cheng Kung University, Tainan 70101, Taiwan}

\author{Neill Lambert}
\affiliation{Theoretical Quantum Physics Laboratory, RIKEN Cluster for Pioneering Research, Wako-shi, Saitama 351-0198, Japan}

\author{Franco Nori}
\affiliation{Theoretical Quantum Physics Laboratory, RIKEN Cluster for Pioneering Research, Wako-shi, Saitama 351-0198, Japan}
\affiliation{Department of Physics, University of Michigan, Ann Arbor, Michigan 48109-1040, USA}

\author{Che-Ming Li}
\email{cmli@mail.ncku.edu.tw}
\affiliation{Department of Engineering Science, National Cheng Kung University, Tainan 70101, Taiwan}
\affiliation{Center for Quantum Frontiers of Research $\&$ Technology, National Cheng Kung University, Tainan 70101, Taiwan}
\affiliation{Center for Quantum Technology, Hsinchu 30013, Taiwan}

\begin{abstract}
Remote state preparation (RSP) enables a sender to remotely prepare the quantum state of a receiver without sending the state itself. Recently, it has been recognized that quantum discord is a necessary resource for RSP. Here, we theoretically and experimentally investigate whether RSP can outperform dynamic classical remote state preparation processes. We show that such classical processes can describe certain RSPs powered by quantum discord. Rather, we argue that a new kind of Einstein-Podolsky-Rosen steering for dynamical processes, called quantum process steering, is the resource required for performing nonclassical RSP. We show how to measure quantum process steering by experimentally realizing nonclassical RSP of photonic quantum systems. Moreover, we demonstrate the transition from classical to quantum RSP. Our results also have applications in realizing genuine quantum RSP for quantum-enabled engineering.
\end{abstract}

\maketitle

\begin{figure}[t]
\includegraphics[width=8.75cm]{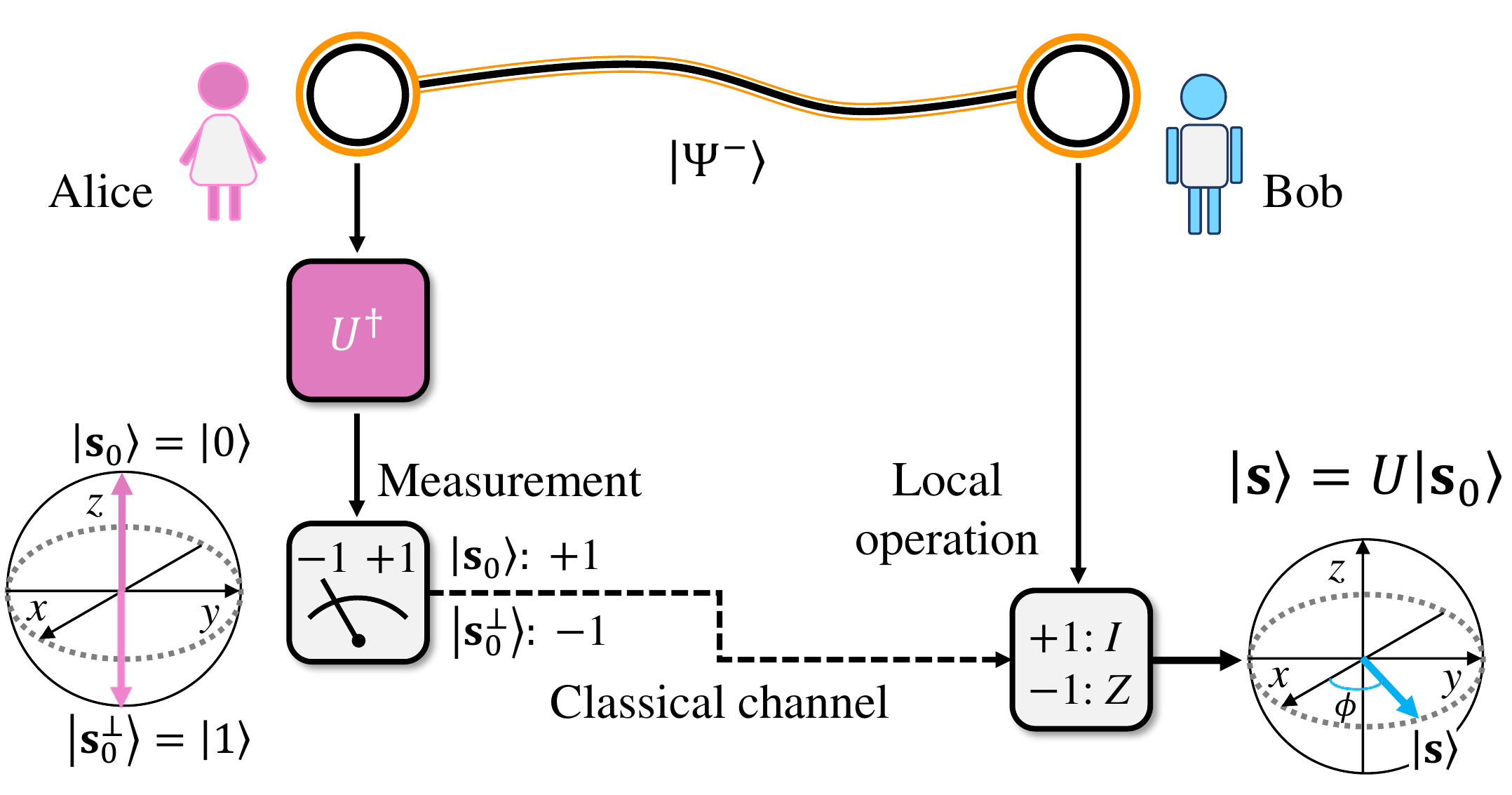}
\caption{Remote state preparation (RSP). To prepare Bob's remote state $\ket{\bold{s}}$ with a desired operation $U$, Alice first performs the operation $U^{\dag}$ and then measures her qubit (part of the state $\ket{\Psi^{-}}$) in the basis $\{\ket{\bold{s}_{0}},\ket{\bold{s}_{0}^{\bot}}\}$. Suppose $\ket{\bold{s}_{0}}=\ket{0}$, $\ket{\bold{s}_{0}^{\bot}}=\ket{1}$, as shown on the Bloch sphere by the Bloch vectors, and $U=R(\phi)$ is a rotation operator. If Alice obtains the result $\ket{0}$ ($\ket{1}$), she sends Bob the message $+1$ ($-1$) via a classical channel to show that his qubit is in the prepared state: $R(\phi)\ket{0}=(\ket{0}+e^{i\phi}\ket{1})/\sqrt{2}$, without correction, denoted as the identity $I$ in the figure (with a correction of the Pauli-$Z$ operation). It is worth noting that, while Bob may know the $U$, the prepared states $\ket{\bold{s}}$ are unknown to him as the $\ket{\bold{s}_{0}}$ is unknown as well. The above procedure equivalently results in a quantum operation $\mathcal{E}_{\text{RSP}}$ which transforms $\ket{\bold{s}_{0}}$ to $\ket{\bold{s}}$, as described in Eq.~(\ref{RSPQO}).}
\label{RSP}
\end{figure}

In the RSP protocol \cite{Pati00,Bennett01}, Alice and Bob initially share an Einstein-Podolsky-Rosen (EPR) pair $\ket{\Psi^{-}}=(\ket{01}-\ket{10})/\sqrt{2}$. Because of rotational symmetry the state $\ket{\Psi^{-}}$ can be rewritten as $\ket{\Psi^{-}}=(U\ket{\bold{s}_{0}}\otimes U\ket{\bold{s}_{0}^{\bot}}-U\ket{\bold{s}_{0}^{\bot}}\otimes U\ket{\bold{s}_{0}})/\sqrt{2}$ \cite{Cabello02}, where $\ket{\bold{s}_{0}}$ and $\ket{\bold{s}_{0}^{\bot}}$ are arbitrary pure states that constitute an orthonormal basis, and $U$ is an arbitrary single qubit unitary operator. If the quantum state, $\ket{\bold{s}}=(\ket{0}+e^{i\phi}\ket{1})/\sqrt{2}$, on the equatorial plane of the Bloch sphere, is the target state envisaged by Alice, she can measure her qubit in the basis $\{U\ket{\bold{s}_{0}},U\ket{\bold{s}_{0}^{\bot}}\}$ to prepare the state $\ket{\bold{s}}=U\ket{\bold{s}_{0}}$ for Bob. Depending on the outcome of Alice's measurement, one classical bit (cbit) is sent from Alice to Bob in order to tell him whether or not he needs to apply a $\pi$ rotation about the $z$ axis for state preparation to be achieved. For example, the state $\ket{\bold{s}}$ can be remotely prepared by using the RSP protocol with $\ket{\bold{s}_{0}}=\ket{0}$ and a rotation operator $U$: $R(\phi)\!=\!(\ket{0}\!\!\bra{0}\!+\!e^{i\phi}\ket{1}\!\!\bra{0}\!+\!\ket{0}\!\!\bra{1}\!-\!e^{i\phi}\ket{1}\!\!\bra{1})\!/\!\sqrt{2}$. See Fig.~\ref{RSP}.

The RSP confers the obvious advantage of remotely manipulating Bob's qubit. In particular, if Bob is unable to locally implement the target operator $U$ but only the simple $\pi$ rotation, Alice can help. Moreover, compared to teleportation \cite{Bennett93}, which requires Bell-state measurements \cite{Pirandola15} for transmitting an unknown qubit, the RSP protocol only needs local measurements on Alice's particle. These two features  makes RSP appealing and well adapted for a range of applications in quantum information and quantum engineering --- from deterministically creating single-photon states \cite{Jeffrey04} to preparing single-photon hybrid entanglement \cite{Barreiro10}, initializing atomic quantum memory \cite{Rosenfeld07}, and assisting teleportation between atomic-ensemble quantum nodes \cite{Bao12}.

Identifying truly quantum RSP, irreproducible by any classical means, is not only significant in its own right, but can be useful for realizing the  power of quantum-enabled technologies in which RSP plays a role \cite{Jeffrey04,Barreiro10,Rosenfeld07,Bao12}. Validation of RSP naturally leads to the following fundamental  question: what are the physical resources actually required to realize nonclassical RSP?

More recently, it was discovered that quantum discord \cite{Ollivier01,Zurek00,Henderson01}, which measures nonclassical correlations between two systems in terms of the difference between two prescribed expressions for mutual information, serves as a resource for RSP \cite{Dakic12}. While maximally-entangled EPR pairs are assumed to be the default states used in the ideal RSP protocol, rather remarkably, states with no entanglement can outperform entangled states in RSP once they possess non-zero geometric quantum discord \cite{Dakic10,Luo10}. 
Importantly, this kind of verification relies on assumptions deduced from  quantum theory. It does not generally hold for practical implementations where experimental imperfections can be described by classical physics. The same assumption is made in the fidelity benchmarks for RSP \cite{Killoran10}, which offer a seminal paradigm for ruling out certain cheating strategies. This issue turns out to hinge on whether it is possible to simulate RSP using classical methods.

Driven by the desire to understand the resources necessary for the nonclassical preparation of quantum remote states, we investigate to what extent RSP can be performed with generic classical dynamical processes. We introduce a model to capture the classical dynamic change to a state which can occur as the result of a remote-state preparation process. The scenario considered in this model generally goes beyond the assumptions made in the existing verification methods \cite{Dakic12,Killoran10}. We show that certain RSPs based on quantum discord can be described by such classical processes. A new kind of EPR steering \cite{Wiseman07,Reid09} is found and proven to be necessary for surpassing such classical preparation of quantum remote states. This enables nonclassicality of experimental implementations to be measured in terms of quantum process steering, which we will introduce below. Experimental demonstration using polarization-correlated photon pairs support our theoretical predictions.

\textit{Theory.}---The whole RSP process can be considered as a quantum operation, $\mathcal{E}_{\text{RSP}}$, implementing the following state transformation:
\begin{equation}
\rho_{\bold{s}}=\mathcal{E}_{\text{RSP}}(\rho_{\bold{s}_{0}}),\label{RSPQO}
\end{equation}
where $\rho_{\bold{s}_{0}}=\ket{\bold{s}_{0}}\!\!\bra{\bold{s}_{0}}$ is the initial state before the RSP process, and $\rho_{\bold{s}}=\ket{\bold{s}}\!\!\bra{\bold{s}}$ is the output state after the process occurs. It is worth emphasizing that RSP is nonclassical since $\mathcal{E}_{\text{RSP}}$ performs a unitary quantum-mechanical transformation $U$. Moreover, $\mathcal{E}_{\text{RSP}}$ comprises all the operations and resources required to implement the RSP protocol. Therefore, any experimental imperfections in these necessary elements can affect the realization of $\mathcal{E}_{\text{RSP}}$. The worst case scenario is when a real-world RSP process may be interpreted as a classical process.

To demonstrate our classical RSP model, we first note that the input state $\ket{\bold{s}_{0}}$ can be represented by $\rho_{\bold{s}_{0}}=(I+\vec{\bold{s}}_{0}\cdot \vec{\sigma})/2$, where $I$ is the identity matrix, $\vec{\bold{s}}_{0}=(s_{01},s_{02},s_{03})$ is a real three-dimensional vector (called the Bloch vector \cite{Nielsen00}, see Fig.~\ref{RSP}) with $|\vec{\bold{s}}_{0}|=1$, and $\vec{\sigma}=(\sigma_{1},\sigma_{2},\sigma_{3})$ consists of $\sigma_{1}=X$, $\sigma_{2}=Y$, and $\sigma_{3}=Z$ Pauli matrices. After a RSP process $\mathcal{E}$, the remote state of Bob's qubit is prepared in the state $\mathcal{E}(\rho_{\bold{s}_{0}})=[\mathcal{E}(I)+\sum_{m=1}^{3}s_{0m}\mathcal{E}(\sigma_{m})]/2$. Explicitly, the observables transform as $\mathcal{E}(I)=\sum_{n=0}^{1}\mathcal{E}(\ket{n}_{mm}\!\!\bra{n})$ and $\mathcal{E}(\sigma_{m})=\sum_{n=0}^{1}v_{nm}\mathcal{E}(\ket{n}_{mm}\!\!\bra{n})$, where $v_{nm}=(-1)^{n}$ and $\ket{n}_{m}$ are the eigenvalues and eigenvectors of $\sigma_{m}$, respectively. As shown above, for a given initial state $\rho_{\bold{s}_{0}}$, the remote state $\mathcal{E}(\rho_{\bold{s}_{0}})$ is determined by the RSP process applied to the eigenstates, $\mathcal{E}(\ket{n}_{mm}\!\!\bra{n})$.

In our model of classical RSP, we assume that the whole process is performed in the absence of EPR pairs. Furthermore, Alice is assumed to be unable to implement any unitary transformations. Instead, the classical RSP consists of the following three steps. First, there exists a recipe which specifies how the eigenstates $\ket{n}_{mm}\!\!\bra{n}$ are described by classical pre-existing states. This step represents Alice identifying which pre-existing state she has. The second step describes how the classical states change during the RSP process. The transition of classical states mimics the unitary transformation in the RSP task. Finally, the final classical states determine the qubit states that Bob has at the end of the classical RSP process.


A classical pre-existing state is a outcome set composed of the pre-existing outcomes corresponding to the observables $\sigma_{m}$ under the assumption of classical realism \cite{Brunner14}, and is described by $v_{\lambda}\equiv(v_{n_{1}1},v_{n_{2}2},v_{n_{3}3})$, where $\lambda=4n_{1}+2n_{2}+n_{3}+1$ for $n_{1},n_{2},n_{3}\in\{0,1\}$ is a unique index determining the set. In a classical RSP, denoted as $\mathcal{E}_{c}$, an initial state, say $\ket{n}_{mm}\!\!\bra{n}$, is interpreted as a pre-existing (realistic) outcome $v_{nm}$ of certain pre-existing states $v_{\lambda}$. For example, the classical states $v_{\lambda}$ (for $\lambda=1,2,3,4$) possess the same state property of $v_{n_{1}1}=v_{01}$. Moreover, $\mathcal{E}_{c}$ causes the $v_{\lambda}$ to classically transit to another classical pre-existing state. After the state transition has occurred, the final classical state, say $v_{\mu}$, indicates a possible remote state $\rho_{\mu}$ for Bob's ``qubit''.

The RSP in the above scenario can be described by the map
\begin{equation}
\mathcal{E}_{c}(\ket{n}_{mm}\!\!\bra{n})=\sum_{\lambda}p(\lambda|v_{nm})\;\tilde{\rho}_{\lambda}\ \ \ \ \forall \ n,m,\label{ec}
\end{equation}
where $p(\lambda|v_{nm})$ is the probability of finding $v_{\lambda}$ conditioned on the pre-existing outcome $v_{nm}$ with $\sum_{\lambda}p(\lambda|v_{nm})=1$, $\tilde{\rho}_{\lambda}=\sum_{\mu}\Omega_{\mu\lambda}\rho_{\mu}$ denotes Bob's possible remote states, and $\Omega_{\mu\lambda}$ is the probability of transition from $v_{\lambda}$ to $v_{\mu}$. The process $\mathcal{E}_{c}$ then classically prepares an arbitrary input state $\rho_{\bold{s}_{0}}$ as $\mathcal{E}_{c}(\rho_{\bold{s}_{0}})=\rho_{\bold{r}_{c}|\bold{s}_{0}}$, where
\begin{equation}
\rho_{\bold{r}_{c}|\bold{s}_{0}}\!\!=\!\frac{1}{2}[\mathcal{E}_{c}(I)+\!\sum_{m,n}s_{0m}v_{nm}\mathcal{E}_{c}(\ket{n}_{mm}\!\!\bra{n})]\ \ \forall \ket{\bold{s}_{0}},\label{rrc}
\end{equation}
with a Bloch vector $\vec{\bold{r}}_{c}$. This process of preparing remote states is classical in the sense that the initial states and their subsequent evolution are both classical, i.e., the essential elements in a dynamical map all follow a local and realistic classical theory.

The resulting state [Eq.~(\ref{rrc})] of the classical map [Eq.~(\ref{ec})] is a generalization of the local hidden state (LHS) model \cite{Wiseman07,Reid09}, including the temporal analogue of the LHS model \cite{Chen14,Chen16}. The LHS model demonstrates how Alice uses a classical strategy to affect the states of Bob's qubit without EPR pairs; the invalidation of this model shows EPR steering of the shared pair. In contrast, Eq.~(\ref{rrc}) describes the relationship between all possible inputs from Alice and outputs for Bob of a classical dynamical process designed to simulate a RSP task. In addition to Alice's measurement, which is only the action taken in a steering scenario, Eq.~(\ref{rrc}) emulates all the necessaties required to realize the RSP protocol, including Alice's manipulation related to $U$, classical communication, and Bob's correction operations.

Compared to the conventional model describing \cite{Wiseman07,Reid09} and quantifying \cite{Skrzypczyk2014,piani2015} EPR steering, the generalized LHS model is stricter because it is a larger class consisting of LHS models and the criterion describing dynamical processes. These constraints make the generalized LHS model effective in examining all possible output states for all $\ket{\bold{s}_{0}}$ in a RSP task. Any $\mathcal{E}$ with outputs that cannot be represented in the form [Eq.~(\ref{rrc})] for any pure inputs implements a nonclassical map, which is called \textit{quantum process steering}. See Supplemental Material \cite{SuppMaterial} for detailed comparisons between this generalized LHS model~[Eq.~(\ref{rrc})] and the standard LHS model.



\begin{figure}
\includegraphics[width=8.7cm]{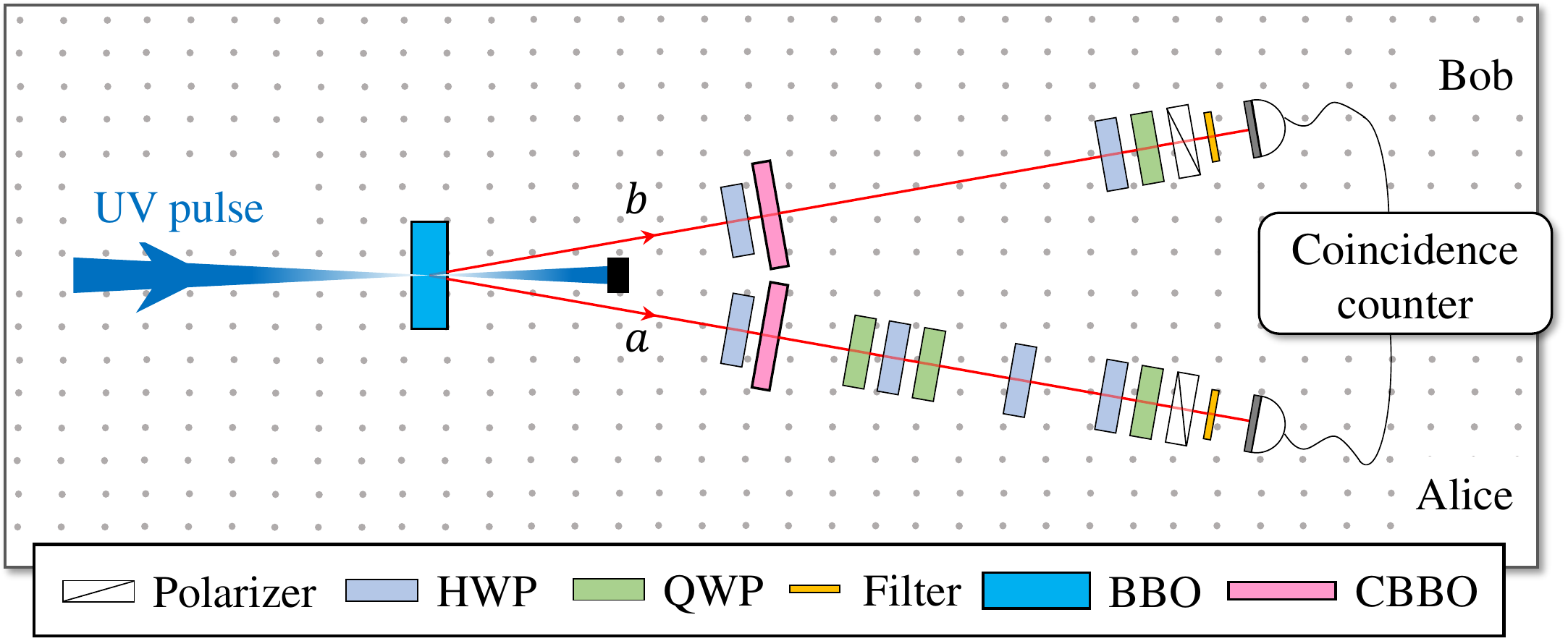}
\caption{Experimental set-up for RSP. Polarization-correlated photon pairs are created via type-II spontaneous parametric down-conversion (SPDC) \cite{Kwiat95} from a $\beta$-barium
borate (BBO) crystal with 2 mm thickness at a wavelength of 780 nm, where the BBO crystal is pumped with a laser beam at 390 nm with a repetition rate of 76 MHz. The longitudinal and spatial walk-off of the photons in modes $a$ and $b$ are compensated by a half-wave plate (HWP) at $90^{\circ}$ and a correction BBO (CBBO) of 1 mm thickness. A combination of two quarter-wave plates (QWP) and one HWP is used to control the phase introduced by optical elements such that the entangled photons are prepared in the state $\ket{\Psi^{-}}$. All photons are filtered by narrow bandwidth filters ($\Delta \lambda \sim3$ nm) and are monitored by silicon avalanche single-photon detectors. Coincidences are recorded by a field-programmable gate-array-based coincidence unit.}
\label{setup}
\end{figure}

\begin{figure}
\includegraphics[width=8.9cm]{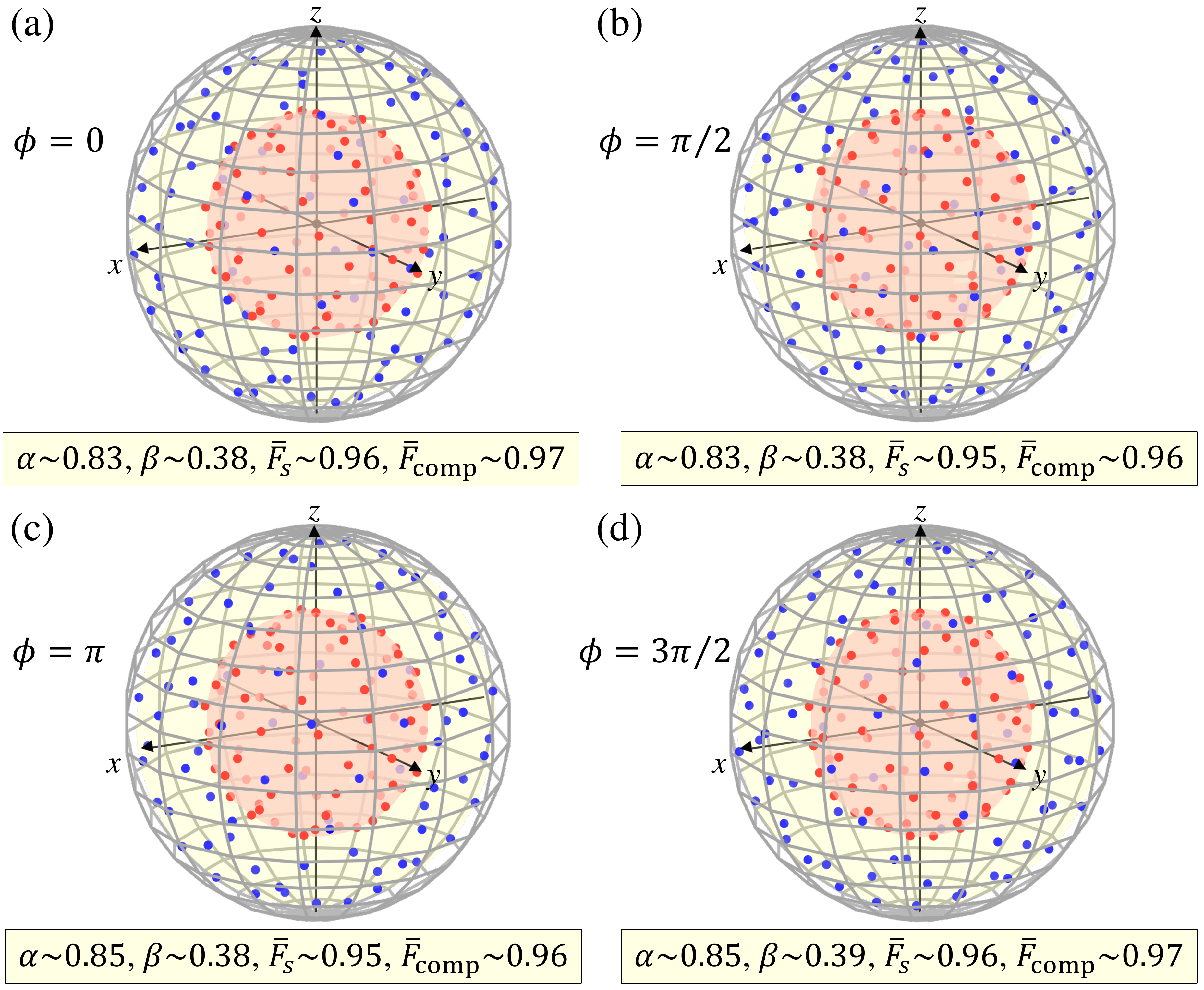}
\caption{Visualization of nonclassical RSPs. We experimentally demonstrated nonclassical RSPs, $\mathcal{E}_{\phi}$, for (a) $\phi=0$, (b) $\phi=\pi/2$, (c) $\phi=\pi$, and (d) $\phi=3\pi/2$. The errors from Poissonian noise in all fidelities are below $0.5\%$. The quantumness of $\mathcal{E}_{\phi}$ can be visualized from the output states $\rho_{\bold{r}|\bold{s}_{0}}$ shown on the Bloch sphere. The $\rho_{\bold{r}|\bold{s}_{0}}$ generated by the ideal preparation $\mathcal{E}^{(\phi)}_{\text{RSP}}$ are on the surface of the Bloch sphere (gray mesh). Conditioned on $98$ randomly chosen pure input states $\rho_{\bold{s}_{0}}$, we observed all the corresponding output states (blue points) of $\mathcal{E}_{\phi}$ having larger Bloch vectors, $|\vec{\bold{r}}|>|\vec{\bold{r}}_{c}|$, compared to the states $\rho_{\bold{r}_{c}|\bold{s}_{0}}$ (red points) derived from the classical RSP $\mathcal{E}_{c}$ with the best process fidelity.}
\label{UT}
\end{figure}

To evaluate how well a given experimental process $\mathcal{E}$ surpasses the generalized LHS model, we compare its output, $\mathcal{E}(\rho_{\bold{s}_{0}})=\rho_{\bold{r}|\bold{s}_{0}}$, with all possible outputs from classical processes, $\rho_{\bold{r}_{c}|\bold{s}_{0}}$, for all pure input states $\rho_{\bold{s}_{0}}$. We illustrate four different methods to measure and (or) identify the capability of $\mathcal{E}$ to exhibit quantum process steering, called \textit{process steerability}.

\noindent (i) Quantum composition $\alpha$:
\begin{equation}
\rho_{\bold{r}|\bold{s}_{0}}=\alpha\rho_{\bold{r}_{Q}|\bold{s}_{0}}+(1-\alpha)\rho_{\bold{r}_{c}|\bold{s}_{0}} \ \ \ \ \forall \ket{\bold{s}_{0}},\label{alpha}
\end{equation}
where $\alpha$ describes the minimum proportion of $\rho_{\bold{r}|\bold{s}_{0}}$ as being the $\rho_{\bold{r}_{Q}|\bold{s}_{0}}$ that cannot be represented by $\rho_{\bold{r}_{c}|\bold{s}_{0}}$.

\noindent (ii) Quantum robustness $\beta$:
\begin{equation}
\frac{\rho_{\bold{r}|\bold{s}_{0}}+\beta\rho_{\text{noise}|\bold{s}_{0}}}{1+\beta}=\rho_{\bold{r}_{c}|\bold{s}_{0}} \ \ \ \ \forall \ket{\bold{s}_{0}},\label{beta}
\end{equation}
where $\beta$ is the minimum amount of the noise $\rho_{\text{noise}|\bold{s}_{0}}$ added such that the $\rho_{\bold{r}|\bold{s}_{0}}$ becomes classical.

\noindent (iii) Average-state-fidelity criterion for quantum RSP:
\begin{equation}
\bar{F}_{s}(\mathcal{E})\!\equiv\!\int \!\!d\bold{s}_{0}\!\bra{\bold{s}_{0}}\!U^{\dag}\!\mathcal{E}(\rho_{\bold{s}_{0}})U\!\ket{\bold{s}_{0}}\; >\;\!\bar{F}_{sc}\;\sim\; 0.79,\label{Fc}
\end{equation}
where $\bar{F}_{sc}\!\equiv\!\max_{\mathcal{E}_{c}}\bar{F}_{s}(\mathcal{E}_{c})$, and the integral is over the uniform measure $d\bold{s}_{0}$ on state space and $\int d\bold{s}_{0}=1$.

\begin{figure*}
\includegraphics[width=18cm]{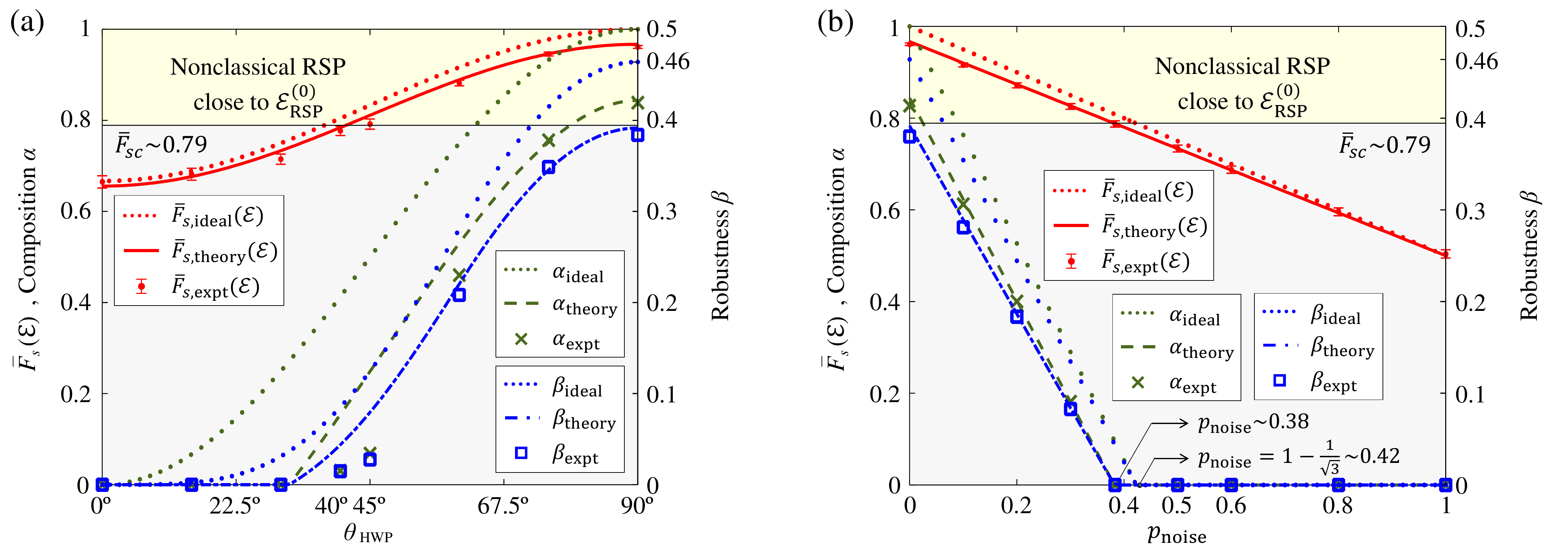}
\caption{Experimental transition from classical to quantum RSP. (a) Experimental RSPs varying with the photon walk-off effect. Given a target RSP, $\mathcal{E}^{(0)}_{\text{RSP}}$, the process steering of experimental RSPs changes with the setting angle ($\theta_{\text{HWP}}$) of the half-wave plate used for compensating the photon walk-off effect (Fig.~\ref{setup}). The walk-off compensation does not exist at $\theta_{\text{HWP}}=0^{\circ}$, which results in a completely mixed state of photon pair. The experimental average state fidelity, $\bar{F}_{s,\text{expt}}(\mathcal{E})$, and process steerability, $\alpha_{\text{expt}}$ and $\beta_{\text{expt}}$, monotonically increase with $\theta_{\text{HWP}}$ and are highly consistent with the theoretical predictions of $\bar{F}_{s,\text{theory}}(\mathcal{E})$, $\alpha_{\text{theory}}$, and $\beta_{\text{theory}}$, where the effect of noise in the photon pairs created has been taken into account \cite{SuppMaterial}. Compared to the theoretical predictions that are calculated using effective noise states, the ideal predictions of $\bar{F}_{s,\text{ideal}}(\mathcal{E})$, $\alpha_{\text{ideal}}$, and $\beta_{\text{ideal}}$ are obtained from perfect photon pairs, $\ket{\Psi^{-}}\!\!\bra{\Psi^{-}}$ \cite{SuppMaterial}. A RSP is nonclassical and close to $\mathcal{E}^{(0)}_{\text{RSP}}$ only when satisfying the criterion~[Eq.~(\ref{Fc})]. For example, the created state at $\theta_{\text{HWP}}=40^{\circ}$ possesses quantum discord \cite{Dakic10} ($\mathcal{D}_{\text{theory}}\sim0.07$, $\mathcal{D}_{\text{expt}}\sim0.08$) \cite{SuppMaterial} and steerability in terms of steerable weight \cite{Skrzypczyk2014} ($SW_{\text{theory}}\sim0.15$, $SW_{\text{expt}}\sim0.02$) \cite{SuppMaterial}, but the resulting RSP is not superior to $\mathcal{E}_{c}$. (b) Experimental RSPs by the Werner states. A variety of polarization-correlated photon pairs in the Werner states $\rho_{W}(p_{\text{noise}})$ are created to realize the RSP protocol. As theoretically shown and experimentally demonstrated, a RSP is nonclassical only when the noise is lower than certain thresholds. Moreover, a $\rho_{W}(p_{\text{noise}})$ which possesses quantum discord can even result in a classical RSP, e.g., ($\mathcal{D}_{\text{theory}}\sim0.11$, $\mathcal{D}_{\text{expt}}\sim0.10$) for $p_{\text{noise}}=0.5$, where the effect of noise in created photon pairs is also taken into account \cite{SuppMaterial}. The ideal values of $\bar{F}_{s,\text{ideal}}(\mathcal{E})$, $\alpha_{\text{ideal}}$, and $\beta_{\text{ideal}}$ are calculated using $\rho_{W}(p_{\text{noise}})$.}
\label{IM}
\end{figure*}

\noindent (iv) Complementary-state-fidelity criterion for quantum RSP:
\begin{equation}
\bar{F}_{\text{comp}}(\mathcal{E})\!\!\equiv\!\!\frac{1}{4}\!\sum_{q=1}^{2}\sum_{n=0}^{1}\!F^{(q)}_{sn}(\mathcal{E})\!\;>\;\!\frac{1}{4}(3\bar{F}_{sc}+1)\sim 0.84,\label{Fcomp}
\end{equation}
where $F^{(q)}_{sn}(\mathcal{E})\equiv \bra{\bold{s}^{(q)}_{0n}}\!U^{\dag}\!\mathcal{E}(\rho_{\bold{s}^{(q)}_{0n}})U\!\ket{\bold{s}^{(q)}_{0n}}$ for $q=1,2$ denote the state fidelities over arbitrary two orthonormal and complementary state bases, $S_{1}=\{\ket{\bold{s}^{(1)}_{00}},\ket{\bold{s}^{(1)}_{01}}\}$ and $S_{2}=\{\ket{\bold{s}^{(2)}_{00}},\{\ket{\bold{s}^{(2)}_{01}}\}$, with $|\langle \bold{s}^{(1)}_{0n}|\bold{s}^{(2)}_{0n'}\rangle|=1/\sqrt{2}$ for $n,n'=0,1$.

In measuring $\alpha$ or $\beta$, or determining the threshold $\bar{F}_{sc}$, the optimization problem with respect to the dynamical maps $\mathcal{E}_{c}$ can be solved by semi-definite programming \cite{SuppMaterial}. The criterion~[Eq.~(\ref{Fcomp})] is based on the fact that, since $\bar{F}_{\text{comp}}(\mathcal{E})\leq (3\bar{F}_{s}(\mathcal{E})+1)/4$ \cite{Hofmann05}, beating the classical limit implies a quantum RSP. While the methods defined in (i)-(iii) can characterize the whole RSP process for all pure input states $\ket{\bold{s}_{0}}$, practical experimental realizations require complete process tomography \cite{Nielsen00}. In contrast, implementing the criterion (iv), Eq.~(\ref{Fcomp}), is efficient and requires only two complementary sets of input states. For example, to identify experimental RSPs close to a target $\mathcal{E}_{\text{RSP}}$ of $U=R(\phi)$, one can follow the RSP protocol and examine the experimental results with the criterion~[Eq.~(\ref{Fcomp})], using $\ket{\bold{s}_{0}}\in S_{n}\!=\!\{\ket{0}_{n+1},\ket{1}_{n+1}\}$, for $n=1,2$, as input states.


We now consider the process steerability of the ideal RSP process for arbitrary $U$. In this ideal case, $\mathcal{E}_{\text{RSP}}$ has a quantum composition of $\alpha=1$, a quantum robustness of $\beta=0.464$, and for the fidelity criteria~[Eqs.~(\ref{Fc}) and~(\ref{Fcomp})], gives $\bar{F}_{s}(\mathcal{E}_{\text{RSP}})=\bar{F}_{\text{comp}}(\mathcal{E}_{\text{RSP}})=1$. These values will decrease when contaminated with noise. For example, when the Werner state $\rho_{W}(p_{\text{noise}})=(1-p_{\text{noise}})\ket{\Psi^{-}}\!\!\bra{\Psi^{-}}+p_{\text{noise}}I\otimes I/4$, is used for RSP to represent noise in the state creation process, the process steerability decreases with increasing noise intensity $p_{\text{noise}}$ and disappears at $p_{\text{noise}}=1-1/\sqrt{3}\sim0.42$, while the underlying Werner state still possesses a non-zero quantum discord.

Process steerability serves as the resource required to yield a higher payoff for nonclassical RSP. The fidelity thresholds in the criteria [Eqs.~(\ref{Fc}) and~(\ref{Fcomp})] are stricter than the fidelity threshold, $\bra{\bold{s}_{0}}\!U^{\dag}\!\mathcal{E}(\rho_{\bold{s}_{0}})U\!\ket{\bold{s}_{0}} =1/2$, for the RSPs enabled by quantum discord \cite{Dakic12}. Furthermore, compared with existing fidelity benchmarks \cite{Killoran10}, our formalism allows, in the classical regime, for imperfections in Alice's results, and removes the assumption that can cause false-positive results in the benchmarks proposed in Ref. \cite{Killoran10}, i.e., when Bob knows the target states in the cheating strategies. Therefore, our fidelity criteria eases the necessary requirements of high state fidelity in the benchmarks.

\textit{Experiment and results.}---In the experiment, we generate photon pairs correlated in the polarization degree of freedom. A schematic diagram of our experimental set-up is shown in Fig.~\ref{setup}. We first implement four different kinds of RSP, denoted as $\mathcal{E}_{\phi}$, while we denote the ideal RSP process as $\mathcal{E}^{(\phi)}_{\text{RSP}}$, which are identified by the choice of single-qubit unitaries, $U=R(\phi)$. The experimental results for the criteria outlined in (i)-(iv), and the concrete comparisons between $\mathcal{E}_{\phi}$, $\mathcal{E}^{(\phi)}_{\text{RSP}}$ and $\mathcal{E}_{c}$ are shown in Fig.~\ref{UT}. The experimental RSPs exhibit quantum process steering and are seen to be close to the target RSP according to the fidelity criteria, which confirms the realizations of nonclassical RSPs.

In Fig.~\ref{IM}(a), we demonstrate how the photon walk-off effect affects the process steerability of the resulting RSPs.  The walk-off effect describes that the ordinary and the extraordinary photons created via type-II SPDC possess different velocities inside the BBO crystal. Such effect makes generated photon pairs separable and the resulting experimental RSP classical. The walk-off effect can be compensated by a HWP and a CBBO in each mode $a$ and $b$ (see Fig.~\ref{setup}). By varying the compensation of the walk-off of the photons from SPDC, the generated photon pairs can contribute to different process steerabilities. Figure~\ref{IM}(b) shows that experimental process steerability varies with the mixedness of the Werner state $\rho_{W}(p_{\text{noise}})$. To experimentally set the noise weight of $p_{\text{noise}}$, the Werner states were generated by appropriately arranging the measurement durations \cite{Lavoie10,Amselem09}.


In Fig.~\ref{IM}(a), we show an example where a photon pair with non-zero quantum discord and EPR steerability results in an RSP that does not satisfy the fidelity criteria.
The experimental results in Fig.~\ref{IM}(b) illustrate further that $\mathcal{E}_{c}$ can even describe certain RSP processes which rely on quantum discord. They are consistent with the theoretical analysis showing that quantum process steering enables quantum RSP.

We have shown that quantum process steering is the necessary resource that enables RSP to surpass classical dynamical methods of preparing remote states. We experimentally demonstrated this new kind of EPR steering, via several criteria, using different types of polarization-correlated photon states. Our demonstration also shows that other quantum characteristics of polarization-correlated photon pairs, such as quantum discord and standard EPR steering, do not guarantee the existence of nonclassical RSP. Our results outline how to achieve genuine quantum RSP, and are of importance for applications where ruling out any classical mimicries of RSP is necessary. They may provide a useful insight into identifying nonclassical processes for other quantum-information tasks, such as universal one-way quantum computing \cite{raussendorf2001,briegel2009,you2007,tanamoto2009,wang2010} or quantum communication in quantum networks \cite{Chou18,Pirker18}.

\textit{Acknowledgement.---}We acknowledge the fruitful discussions with He Lu. This work was partially supported by the Ministry of Science and Technology, Taiwan, under Grant No. MOST 107-2628-M-006-001-MY4. F.N. is supported in part by: NTT Research, Army Research Office (ARO) (Grant No. W911NF-18-1-0358), Japan Science and Technology Agency (JST) (via the CREST Grant No. JPMJCR1676), Japan Society for the Promotion of Science (JSPS) (via the KAKENHI Grant No. JP20H00134, and the grant JSPS-RFBR Grant No. JPJSBP120194828), and the Grant No. FQXi-IAF19-06 from the Foundational Questions Institute Fund (FQXi), a donor advised fund of the Silicon Valley Community Foundation. N.L. is partially supported by JST PRESTO through Grant No. JPMJPR18GC.


\section*{Supplementary Information}

\section{Comparisons between the generalized LHS model and the standard LHS model}

The local hidden state (LHS) model \cite{Wiseman07,Reid09} describes the state of Bob's qubit conditioned on Alice's measurement:
\begin{equation}
\rho_{nm}=\sum_{\lambda}p(\lambda|v_{nm})\rho_{\lambda} \ \ \ \ \ \ \ \forall n,m,\label{r}
\end{equation}
where $v_{nm}$ denotes the $n$th measurement outcome of Alice's $m$th measurement. Alternatively, this can be represented by unnormalized states and rephrased as
 \begin{equation}
\rho'_{nm}=\sum_{\lambda}p(v_{nm}|\lambda)\rho'_{\lambda} \ \ \ \ \ \ \ \forall n,m,\label{rp}
\end{equation}
where $\rho'_{nm}=p(v_{nm})\rho_{nm}$ and $\rho'_{\lambda}=p(\lambda)\rho_{\lambda}$, and $p(v_{nm}|\lambda)$ is the probability of finding the outcome $v_{nm}$ conditioned on the classical state $v_{\lambda}$. The relation of $p(v_{nm})p(\lambda|v_{nm})=p(\lambda)p(v_{nm}|\lambda)$ has be used to derive Eq.~(\ref{rp}) from Eq.~(\ref{r}). Given Alice's choice of measurement $m$ and outcome $v_{nm}$, any assemblage $\{\rho'_{nm}\}_{nm}$ of unnormalized states which Alice's measurement affects Bob's particle into which cannot be represented in the form~(\ref{rp}) is called steerable \cite{Wiseman07,Reid09}.

To discuss the difference between our classical model and the LHS model, we first note that, given the input state $\ket{n}_{mm}\!\!\bra{n}$, a classical RSP, $\mathcal{E}_{c}$, generates an output described by Eq.~(2) in the main text:
\begin{equation}
\mathcal{E}_{c}(\ket{n}_{mm}\!\!\bra{n})=\sum_{\lambda}\sum_{\mu}p(\lambda|v_{nm})\Omega_{\mu\lambda}\rho_{\mu}\ \ \ \ \forall \ n,m,\label{main2}
\end{equation}
which can be written as
\begin{equation}
\mathcal{E}_{c}(\ket{n}_{mm}\!\!\bra{n})=\sum_{\lambda}p(\lambda|v_{nm})\rho_{\lambda}\ \ \ \ \forall \ n,m, \label{main22}
\end{equation}
under the condition of $\Omega_{\mu\lambda}=\delta_{\mu\lambda}$. When comparing Eq.~(\ref{r}) and Eq.~(\ref{main22}) we summarize the following two points:\\

\noindent (i) While they are of the same form and Eq.~(\ref{main22}) can be considered as a LHS model, $\mathcal{E}_{c}$ cannot be fully defined by considering only some of the output states $\mathcal{E}_{c}(\ket{n}_{mm}\!\!\bra{n})$ under the inputs $\ket{n}_{m}$ in the LHS model via Eq.~(\ref{main22}). Rather, since the $\mathcal{E}_{c}$ is a dynamical process, it can be completely defined only when output states for all the pure input states are taken into account. Note that, it is unnecessary to include mixed input states because they can be generated by incoherently mixing pure states.

\noindent (ii) From the task-oriented viewpoint, any assemblage $\{\rho_{\bold{r}|\bold{s}_{0}}\}$ of output states derived from an experimental RSP process which cannot described by
\begin{equation}
\rho_{\bold{r}_{c}|\bold{s}_{0}}\!\!=\!\frac{1}{2}[\mathcal{E}_{c}(I)+\!\sum_{m,n}s_{0m}v_{nm}\mathcal{E}_{c}(\ket{n}_{mm}\!\!\bra{n})]\ \ \forall \ket{\bold{s}_{0}},\label{rrc}
\end{equation}
as shown in Eq.~(3) in the main text, is called process steerable, which corresponds to the term introduced in the main text: quantum process steering.

For the above reasons, the resulting state (\ref{rrc}) of the classical map $\mathcal{E}_{c}$ therefore is a generalized LHS model. Such a generalization makes identifying and quantifying quantum process steering rather different from the methods for quantitatively characterizing EPR steering. See the next section for further detailed discussions.

The above analysis, comparisons and conclusion are applicable to the temporal LHS model for temporal steering \cite{Chen14,Chen16}. In the temporal scenario, Alice performs a measurement on a single system in a certain initial state at time $t=0$. After the measurement, the initial state becomes, say $\ket{n}_{mm}\!\!\bra{n}$. Then this state is sent into a quantum channel $\mathcal{E}$ for a time $t$. At time $t$, Bob receives the system of the state $\mathcal{E}(\ket{n}_{mm}\!\!\bra{n})=\rho_{nm}(t)$. Following the un-normalized assemblage \cite{Wiseman07,Reid09} in the standard LHS model, the un-normalized states in temporal scenario is defined by $\rho^{T}_{nm}=p(v_{nm})\rho_{nm}(t)$, where the superscript $T$ reminds one that the assemblage $\{\rho^{T}_{nm}\}_{nm}$ is for temporal steering. In the temporal LHS model, an unsteerable assemblage is defined as
 \begin{equation}
\rho^{T}_{nm}=\sum_{\lambda}p(v_{nm}|\lambda)\rho^{T}_{\lambda} \ \ \ \ \ \ \ \forall n,m,\label{rpt}
\end{equation}
where $\rho^{T}_{\lambda}=p(\lambda)\rho_{\lambda}(t)$ and $\rho_{\lambda}(t)$ is created from a source which determines the possible correlations between Alice's (at time $t=0$) and Bob's measurement results (at the time $t$) under classical realism. By using the same method as shown above, Eq.~(\ref{rpt}) can be rephrased as
\begin{equation}
\rho_{nm}(t)=\sum_{\lambda}p(\lambda|v_{nm})\rho_{\lambda}(t) \ \ \ \ \ \ \ \forall n,m.\label{rt}
\end{equation}
Still, when comparing Eq.~(\ref{main22}) to Eq.~(\ref{rt}), $\mathcal{E}_{c}$ cannot be fully defined by considering the finite output states $\mathcal{E}_{c}(\ket{n}_{mm}\!\!\bra{n})$ under the inputs $\ket{n}_{mm}$ in the temporal LHS model via Eq.~(\ref{main22}). The $\mathcal{E}_{c}$ can be defined only when the output states, $\mathcal{E}_{c}(\rho_{\bold{s}_{0}})=\rho_{\bold{r}_{c}|\bold{s}_{0}}(t)$, for all the pure input states are considered, where
\begin{equation}
\rho_{\bold{r}_{c}|\bold{s}_{0}}(t)\!\!=\!\frac{1}{2}[\mathcal{E}_{c}(I)+\!\!\!\sum_{m,n,\lambda}\!\!\!s_{0m}v_{nm}\mathcal{E}_{c}(\ket{n}_{mm}\!\!\bra{n})]\ \ \forall \ket{\bold{s}_{0}}.\label{rrct}
\end{equation}

\section{Quantum composition, quantum robustness, and the fidelity criteria}
In this section, we will introduce how to calculate quantum composition $\alpha$, quantum robustness $\beta$, and the fidelity bound $\bar{F}_{sc}$ in the fidelity criteria. The methods introduced in the main text to measure and (or) identify process steerability of an experimental process $\mathcal{E}$ are summarized as follows:\\

\noindent (i) Quantum composition $\alpha$. For all possible pure states $\ket{\bold{s}_{0}}$ (i.e., $|\vec{\bold{s}}_{0}|=1$), the corresponding prepared remote states, $\mathcal{E}(\rho_{\bold{s}_{0}})=\rho_{\bold{r}|\bold{s}_{0}}$, can be decomposed as
\begin{equation}
\rho_{\bold{r}|\bold{s}_{0}}=\alpha\rho_{\bold{r}_{Q}|\bold{s}_{0}}+(1-\alpha)\rho_{\bold{r}_{c}|\bold{s}_{0}} \ \ \ \ \forall \ket{\bold{s}_{0}},\label{alpha}
\end{equation}
where the quantum composition $\alpha$ describes the minimum proportion of $\rho_{\bold{r}|\bold{s}_{0}}$ that cannot be represented by the classically prepared states $\rho_{\bold{r}_{c}|\bold{s}_{0}}$. An ideal RSP shows that $\rho_{\bold{r}|\bold{s}_{0}}$ is fully composed of the $\rho_{\bold{r}_{Q}|\bold{s}_{0}}$ that cannot be represented by $\rho_{\bold{r}_{c}|\bold{s}_{0}}$, i.e., $\alpha=1$.

\noindent (ii) Quantum robustness $\beta$. In this case the prepared states can be written as
\begin{equation}
\frac{\rho_{\bold{r}|\bold{s}_{0}}+\beta\rho_{\text{noise}|\bold{s}_{0}}}{1+\beta}=\rho_{\bold{r}_{c}|\bold{s}_{0}} \ \ \ \ \forall \ket{\bold{s}_{0}},\label{beta}
\end{equation}
where the quantum robustness $\beta$ is the minimum amount of the noise $\rho_{\text{noise}|\bold{s}_{0}}$ added such that the $\rho_{\bold{r}|\bold{s}_{0}}$ becomes a classically prepared remote state. When $\beta=0$, the experimental RSP is already classical. Increasing $\beta$ increases the difference between $\rho_{\bold{r}|\bold{s}_{0}}$ and $\rho_{\bold{r}_{c}|\bold{s}_{0}}$.

\noindent (iii) Average state fidelity. The $\mathcal{E}$ is identified as being faithful to show quantum process steering if the state fidelity of $\mathcal{E}(\rho_{\bold{s}_{0}})$ and $\mathcal{E}_{\text{RSP}}(\rho_{\bold{s}_{0}})$: $F_{s,\bold{s}_{0}}(\mathcal{E})\equiv\bra{\bold{s}_{0}}\!U^{\dag}\!\mathcal{E}(\rho_{\bold{s}_{0}})U\!\ket{\bold{s}_{0}}$, averaged over all possible input pure states is larger than the maximum averaged state fidelity derived from classical RSP, $\bar{F}_{sc}\!\equiv\!\max_{\mathcal{E}_{c}}\bar{F}_{s}(\mathcal{E}_{c})$, i.e.,
\begin{equation}
\bar{F}_{s}(\mathcal{E})\!\equiv\!\int \!d\bold{s}_{0} F_{s,\bold{s}_{0}}(\mathcal{E}) >\!\bar{F}_{sc}\sim 0.79,\label{Fc}
\end{equation}
where the integral is over the uniform measure $d\bold{s}_{0}$ on state space and $\int d\bold{s}_{0}=1$.

\noindent (iv) Complementary state fidelity. Suppose that $S_{1}=\{\ket{\bold{s}^{(1)}_{00}},\ket{\bold{s}^{(1)}_{01}}\}$ and $S_{2}=\{\ket{\bold{s}^{(2)}_{00}},\{\ket{\bold{s}^{(2)}_{01}}\}$ are two orthonormal and complementary state bases, where $|\langle \bold{s}^{(1)}_{0n}|\bold{s}^{(2)}_{0n'}\rangle|=1/\sqrt{2}$ for $n,n'=0,1$. Let us consider the state fidelities $F^{(q)}_{sn}(\mathcal{E})\equiv \bra{\bold{s}^{(q)}_{0n}}\!U^{\dag}\!\mathcal{E}(\rho_{\bold{s}^{(q)}_{0n}})U\!\ket{\bold{s}^{(q)}_{0n}}$ for $q=1,2$. The $\mathcal{E}$ is faithful and possess quantum process steerability if the complementary state fidelity $\bar{F}_{\text{comp}}(\mathcal{E})$ \cite{Hofmann05} beating the classical limit implies quantum RSP:
\begin{equation}
\bar{F}_{\text{comp}}(\mathcal{E})\!\!\equiv\!\!\frac{1}{4}\!\sum_{q=1}^{2}\sum_{n=0}^{1}\!F^{(q)}_{sn}(\mathcal{E})\!>\!\bar{F}_{\text{comp},c}\equiv\frac{1}{4}(3\bar{F}_{sc}+1)\sim 0.84.\label{Fcomp}
\end{equation}

In measuring the $\alpha$, $\beta$, $\bar{F}_{s}$ and determining the threshold $\bar{F}_{sc}$ and $\bar{F}_{\text{comp},c}$, it is necessary to consider the contributions from all pure states $\ket{\bold{s}_{0}}$ in the respective computational tasks. To solve such questions, we first note that Eq.~(\ref{alpha}) and Eq.~(\ref{beta}) can be rephrased as
\begin{equation}
\mathcal{E}(\rho_{\bold{s}_{0}})=\alpha\mathcal{E}_{Q}(\rho_{\bold{s}_{0}})+(1-\alpha)\mathcal{E}_{c}(\rho_{\bold{s}_{0}})\label{alphaE}
\end{equation}
and
\begin{equation}
\frac{\mathcal{E}(\rho_{\bold{s}_{0}})+\beta\mathcal{E}_{\text{noise}}(\rho_{\bold{s}_{0}})}{1+\beta}=\mathcal{E}_{c}(\rho_{\bold{s}_{0}}),\label{betaE}
\end{equation}
respectively, where $\mathcal{E}_{Q}(\rho_{\bold{s}_{0}})=\rho_{\bold{r}_{Q}|\bold{s}_{0}}$ and $\mathcal{E}_{\text{noise}}(\rho_{\bold{s}_{0}})=\rho_{\text{noise}|\bold{s}_{0}}$. Then the optimization problems with respect to states become to the optimization of $\alpha$ and $\beta$ with respect to dynamical maps \cite{Hsieh17}. Moreover, it has been shown that $\bar{F}_{s}(\mathcal{E})=(2F_{\mathcal{E}}+1)/3$ \cite{Gilchrist05,Hofmann05}, implying that the $\bar{F}_{s}$ can be derived from $F_{\mathcal{E}}$, i.e., the process fidelity of $\mathcal{E}$ and $\mathcal{E}_{\text{RSP}}$. Equivalently, the criterion~(\ref{Fc}) becomes $F_{\mathcal{E}}>F_{\mathcal{E}_{c}}$, where the $F_{\mathcal{E}_{c}}$ is obtained by considering the best similarity between $\mathcal{E}_{c}$ and $\mathcal{E}_{\text{RSP}}$. Finally, as $\mathcal{E}$ is experimentally measured, $\alpha$, and $\beta$, $\bar{F}_{sc}$ can therefore be determined by semi-definite programming (SDP) \cite{Lofberg,sdpsolver} and the $\bar{F}_{\text{comp},c}$ can be calculated through $\bar{F}_{sc}$.

In practical experiments, the experimental process $\mathcal{E}$ can be fully described by using the process tomography algorithm \cite{Nielsen00,Chuang97} and then be represented by a positive Hermitian matrix, called process matrix, $\chi_{\text{expt}}$. The process matrix $\chi_{\text{expt}}$ is constructed by the specific output states $\mathcal{E}(\ket{n}_{mm}\!\!\bra{n})$ which are the output states of the eigenstates of three Pauli matrices. Considering the output states $\mathcal{E}(\ket{n}_{mm}\!\!\bra{n})$ from a classical RSP $\mathcal{E}_{c}$, we can use process tomography algorithm to get a process matrix $\chi_{c}$ for the classical RSP $\mathcal{E}_{c}$. Thus, Eq.~(\ref{alphaE}) and Eq.~(\ref{betaE}) can be rephrased as
\begin{equation}
\chi_{\text{expt}}=\alpha\chi_{Q}+(1-\alpha)\chi_{c}\label{alphachi}
\end{equation}
and
\begin{equation}
\frac{\chi_{\text{expt}}+\beta\chi_{\text{noise}}}{1+\beta}=\chi_{c},\label{betachi}
\end{equation}
respectively, and the $F_{\mathcal{E}_{c}}$ can be represented as $F_{\mathcal{E}_{c}}=\max_{\chi_{c}}[\text{tr}(\chi_{c}\chi_{\text{RSP}})]$, where $\chi_{\text{RSP}}$ is the process matrix of $\mathcal{E}_{\text{RSP}}$ and $\chi_{Q}$ and $\chi_{\text{noise}}$ are the process matrices of $\mathcal{E}_{Q}$ and $\mathcal{E}_{\text{noise}}$ respectively. The process matrices $\chi_{\text{expt}}$ and $\chi_{c}$ can be used to solve the optimization problems of the optimization of $\alpha$, $\beta$, and $\bar{F}_{sc}$. To calculate $\alpha$, $\beta$, and $\bar{F}_{sc}$ via SDP, we let the processes in SDP satisfy the definition of process matrices or density matrices by the following constraint. A process matrix $\chi$ and a density matrix $\rho$ must be positive semi-definite, i.e., $\chi\geq0$ and $\rho\geq0$. Since a process matrix is constructed of the output states $\mathcal{E}(\ket{n}_{mm}\!\!\bra{n})$ and the output states of the $\mathcal{E}_{c}$ for classical RSP model is determined by Bob's possible remote states $\tilde{\rho}_{\lambda}$ in Eq.~(2) in the main text, the process matrix $\chi_{c}$ must satisfy $\tilde{\rho}_{\lambda}\geq 0, \forall \lambda$.

According to Eq.~(\ref{alphachi}), by representing $(1-\alpha)\chi_{c}$ as a unnormalized process matrix $\tilde{\chi}_{c}=(1-\alpha)\chi_{c}$, quantum composition $\alpha$ can be obtained by minimizing the following quantity via SDP with MATLAB \cite{Lofberg,sdpsolver}: $\alpha=\min_{\tilde{\chi}_{c}}[1-\text{tr}(\tilde{\chi}_{c})]$, under the following constraints:
\begin{equation}
\begin{split}
\tilde{\chi}_{c}\geq 0;\\
\tilde{\rho}_{\lambda}\geq 0, \forall \lambda;\\
\chi_{\text{expt}}-\tilde{\chi}_{c}\geq 0.\\
\end{split} \nonumber
\end{equation}
The first constraint ensures that the $\tilde{\chi}_{c}$ must be positive semi-definite since it is a process matrix. The second constraint ensures that all possible remote states in classical RSP model are positive semi-definite. The third constraint makes the process matrix of $\mathcal{E}_{Q}$ satisfy the definition of process matrices which need to be positive semi-definite.

As the method used to solve $\alpha$, to calculate the quantum robustness $\beta$, we define $\tilde{\chi}_{c}=(1+\beta)\chi_{c}$. Thus, according to Eq.~(\ref{betachi}), the quantum robustness $\beta$ of $\chi_{\text{expt}}$ can be obtained by using SDP to solve $\beta=\min_{\tilde{\chi}_{c}}[\text{tr}(\tilde{\chi}_{c})-1]$. The constraints for solving $\beta$ via SDP are
\begin{equation}
\begin{split}
\tilde{\chi}_{c}\geq 0;\\
\tilde{\rho}_{\lambda}\geq 0, \forall \lambda;\\
\tilde{\chi}_{c}-\chi_{\text{expt}}\geq 0;\\
\text{tr}(\tilde{\chi}_{c})\geq 1.
\end{split} \nonumber
\end{equation}
The first and second constraints are the same as the constraints used in calculating $\alpha$. The third constraint ensures that $\beta\geq0$. The fourth constraint is the condition of positive semidefiniteness for $\beta\chi_{\text{noise}}=(1+\beta)\chi_{c}-\chi_{\text{expt}}$ in Eq.~(\ref{betachi}).

With the process matrix $\chi_{\text{RSP}}$ of $\mathcal{E}_{\text{RSP}}$, the fidelity bounds $F_{\mathcal{E}_{c}}\equiv\max_{\tilde{\chi}_{c}}[\text{tr}(\tilde{\chi}_{c}\chi_{\text{RSP}})]$ can be solved under the constraints
\begin{equation}
\begin{split}
\tilde{\chi}_{c}\geq 0;\\
\tilde{\rho}_{\lambda}\geq 0, \forall \lambda;\\
\text{tr}(\tilde{\chi}_{c})= 1,
\end{split} \nonumber
\end{equation}
which make the $\tilde{\chi}_{c}$ be a normalized matrix to calculate fidelity. While we get the fidelity bound $F_{\mathcal{E}_{c}}$, the average-state-fidelity bound $\bar{F}_{sc}$ and the complementary-state-fidelity bound $\bar{F}_{\text{comp},c}$ can be calculated through $\bar{F}_{sc}=(2F_{\mathcal{E}_{c}}+1)/3$ and $\bar{F}_{\text{comp},c}=(3\bar{F}_{sc}+1)/4$.

\section{Experimental photon pairs}


The experimental set-up for generating polarization-correlated photon pairs to achieve certain RSP is shown in Fig.~2 in the main text. A type-II $\beta$-barium borate (BBO) crystal with 2 mm thickness is pumped by ultraviolet laser pulses with a central wavelength of 390 nm, pulse duration of 135 fs, repetition rate of 76 MHz, and power of 250 mW to generate polarization-correlated photon pairs at a central wavelength of 780 nm via spontaneous parametric down-conversion (SPDC) process \cite{Kwiat95}. The photon pairs then pass through birefringent compensators, i.e., two half-wave plates (HWP) and two correction BBO (CBBO) crystals with 1 mm thickness, to eliminate the longitudinal and spatial walk-off effects. Note that the setting angle of the HWPs in both mode $a$ and mode $b$ are $90^{\circ}$ under ideal compensation circumstance, i.e., $\theta_{\text{HWP},a}=\theta_{\text{HWP},b}=\theta_{\text{HWP}}=90^{\circ}$. Subsequently, we utilize a phase controller consisting of two quarter-wave plates (QWP) and a HWP to obtain the desired state, $\ket{\Psi^{-}}$. Finally, the photon pairs are filtered by narrow-band filters with 3 nm bandwidth and are detected by silicon avalanche single-photon detectors. 

The resulting state of the created photon pairs, denoted as $\rho_{\text{expt}}$, can be experimentally characterized via state tomography using a pair of polarization analyzers; each consisting of a QWP, a HWP, and a polarizer, in mode $a$ and mode $b$, respectively. The measured density matrix of the created photon pairs after the designed compensation ($\theta_{\text{HWP}}=90^{\circ}$) is of the following form:
\begin{widetext}
\begin{equation}
\rho_{\text{expt}}(90^{\circ})=
\left[
\begin{matrix}
0.02 & -0.01+0.02i & -0.01-0.01i & 0.01i\\
-0.01-0.02i & 0.49 & -0.45+0.03i & 0.01+0.01i\\
-0.01+0.01i & -0.45-0.03i & 0.48 & 0.01-0.02i\\
-0.01i & 0.01-0.01i & 0.01+0.02i & 0.01\\
\end{matrix}
\right],
\label{rho_90}
\end{equation}
\end{widetext}
where the fidelity of $\rho_{\text{expt}}(90^{\circ})$ and
the target state, $\ket{\Psi^{-}}\!\bra{\Psi^{-}}$, is $F=\text{tr}\left[ \rho_{\text{expt}}(90^{\circ})\ket{\Psi^{-}}\!\bra{\Psi^{-}} \right]\sim 0.94$ in our experiment. The maximum-likelihood technique \cite{James01} has been used to make the desnsity matrix $\rho_{\text{expt}}(90^{\circ})$ reasonable.

The created entangled photon pairs of the state $\rho_{\text{expt}}(90^{\circ})$ is used to perform nonclassical RSPs, $\mathcal{E}_{\phi}$, for (a) $\phi=0$, (b) $\phi=\pi/2$, (c) $\phi=\pi$, and (d) $\phi=3\pi/2$ shown in Fig.~3 in the main text. Moreover, $\rho_{\text{expt}}(90^{\circ})$ is also used to demonstrate $\mathcal{E}_{0}$ in the case of the desired compensation ($\theta_{\text{HWP}}=90^{\circ}$) in Fig.~4(a) and in the case of Werner state at $p_{\text{noise}}=0$ in Fig.~4(b) in the main text.


\section{Noise model for experimental photon pairs}


As shown in Eq.~(\ref{rho_90}), there exist undesired components in the created photon pairs, mainly due to the imperfection of BBO crystals, the higher-order terms of down-converted photons, the imperfection of optical components, the accidental coincidences, and the imperfect detection efficiency, which cannot be corrected by compensating the photon walk-off effect. In other words, despite the use of the HWPs and the CBBO crystals for compensating the photon walk-off effect, the polarization state of the created photon pairs is not the ideal state $\ket{\Psi^-}$ in our experiment. To quantitatively describe the resulting noise and imperfection that are present in our experiment, we effectively represent the state in terms of an ideal state $\ket{\Psi^{-}}\!\bra{\Psi^{-}}$ mixed with white noise. Therefore, after the compensation of the photon walk-off effect, the created photon pairs in our noise model is described by
\begin{equation}
\rho_\text{noise}(90^{\circ})=(1-p)\ket{\Psi^{-}}\!\bra{\Psi^{-}}+p\frac{I\otimes I}{4},\ \ \ \ \ \ \ 0\leq p \leq 1,
\label{mix_white}
\end{equation}
where $p$ denotes the intensity of white noise in created photon pairs. The $p$ in our experiment can be estmated by maximizing the fidelity between $\rho_\text{noise}(90^{\circ})$ and the experimentally created photon pairs, $\rho_{\text{expt}}(90^{\circ})$ (\ref{rho_90}). The fidelity between them is defined as
\begin{equation}
\mathcal{F}=\text{tr}\sqrt{\rho_\text{noise}(90^{\circ})^{\frac{1}{2}}\rho_{\text{expt}}(90^{\circ})\rho_\text{noise}(90^{\circ})^{\frac{1}{2}}}. \nonumber
\end{equation}
After performing the maximization task, the fidelity between $\rho_\text{noise}(90^{\circ})$ and $\rho_{\text{expt}}(90^{\circ})$ is $\mathcal{F} \sim 0.99$, and the estimated noise intensity is $p=0.06$. Therefore, the experimental photon pairs in our noise model is
\begin{equation}
\rho_\text{noise}(90^{\circ})=0.94\ket{\Psi^{-}}\!\bra{\Psi^{-}}+0.06\frac{I\otimes I}{4}.
\label{noise006}
\end{equation}

We utilize the effective noise state $\rho_\text{noise}(90^{\circ})$ to calculate the theoretical values of $\alpha_{\text{theory}}$, $\beta_{\text{theory}}$, and $\bar{F}_{s,\text{theory}}(\mathcal{E})$ at $\theta_{\text{HWP}}=90^{\circ}$ in Fig.~4(a) in the main text. Furthermore, as shown in Fig.~4(b) in the main text, the Werner state in our noise model is described as 
\begin{equation}
\rho_{W,\text{noise}}(p_{\text{noise}})=(1-p_{\text{noise}})\rho_\text{noise}(90^{\circ})+p_{\text{noise}}\frac{I\otimes I}{4},
\label{werner}
\end{equation}  
where $0\leq p_{\text{noise}} \leq 1$ is the noise weight in Werner state. We utilize $\rho_{W,\text{noise}}(p_{\text{noise}})$ to estimate the theoretical values of $\alpha_{\text{theory}}$, $\beta_{\text{theory}}$, $\bar{F}_{s,\text{theory}}(\mathcal{E})$, and quantum discord in Fig.~4(b) in the main text. It is worth noting that the experimental results are highly consistent with the theoretical values. It can be inferred that $\rho_{W,\text{noise}}(p_{\text{noise}})$ is a good estimation of the created photon pairs in Werner state varying with the noise weight.

\section{Compensation of the photon walk-off effect}

While a pumped laser beam enters the BBO crystal, it is partially converted into a pair of mutually orthogonally polarized photons, the ordinary and extraordinary beams which are respectively horizontally and vertically polarized. The two-photon state at the output surface of the BBO crystal can be described as \cite{Rubin94}
\begin{widetext}
\begin{equation}
\ket{\Psi}=\eta\int \text{A}(\omega_\text{p})\text{d}\omega_\text{p}\int \text{d}^3\textbf{\textit{k}}_\text{o} \int\text{d}^3\textbf{\textit{k}}_\text{e}\delta(\omega_\text{p}-\omega_\text{o}-\omega_\text{e})\int_{0}^{L}\text{d}ze^{i(\textbf{\textit{k}}_\text{p}-\textbf{\textit{k}}_\text{o}-\textbf{\textit{k}}_\text{e})z}a_{\textbf{\textit{k}}_\text{o}}^{\dagger}a_{\textbf{\textit{k}}_\text{e}}^{\dagger}\ket{\text{vac}},\label{SPDC}
\end{equation}\
\end{widetext}
where the $z$ direction is assumed to be parallel to the pump beam, $\eta$ is the constant comparing with SPDC efficient; $\omega_\text{p}$, $\omega_\text{o}$ and $\omega_\text{e}$ are the frequencies of the pump, ordinary, and extraordinary beams with the wave vectors $\textbf{\textit{k}}_\text{p}$, $\textbf{\textit{k}}_\text{o}$, and $\textbf{\textit{k}}_\text{e}$, respectively; $\text{A}(\omega_\text{p})$ is the spectral profile of the pump pulse; $\delta$-function $\delta(\omega_\text{p}-\omega_\text{o}-\omega_\text{e})$ shows the energy conservation in SPDC; $L=2$ mm is the length of the BBO crystal; $a_{\textbf{\textit{k}}_\text{o}}^{\dagger}$ and $a_{\textbf{\textit{k}}_\text{e}}^{\dagger}$ are the creation operators for the ordinary and extraordinary beams respectively; $\ket{\text{vac}}$ represents the vacuum state. Since the detectors are in the modes $a$ and $b$ and narrow bandwidth filters are added in front of the detectors to select frequencies $\omega_\text{o}=\omega_\text{e}=\omega_\text{p}/2$, Eq.~(\ref{SPDC}) can be simplified as 
\begin{equation}
\ket{\Psi'}=\eta'(a_{\text{0}}^{\dagger}b_{\text{1}}^{\dagger}+e^{i\phi}a_{\text{1}}^{\dagger}b_{\text{0}}^{\dagger})\ket{\text{vac}},\nonumber 
\end{equation}
where $\eta'$ is the normalization constant, and $a_{i}^{\dagger}$, $b_{i}^{\dagger}$ are the creation operators of photon with polarized state $\ket{i}$ for spatial mode $a$, $b$ respectively.
After the normalization, the polarized state of the generated photon pairs is $\ket{\psi}=(\ket{01}+e^{i\phi}\ket{10})/\sqrt{2}$, where 0 and 1 indicate horizontal and vertical polarizations. Since the BBO crystal is a birefringent crystal and the ordinary and extraordinary beams have orthogonal polarization, a longitudinal and spatial walk-off between the ordinary and extraordinary beams is induced by the BBO crystal \cite{Poh_thesis,Lee16}. We assume that SPDC occurs in the center of the BBO crystal, $z=L/2$, the walk-off effect makes horizontally polarized photons be delayed by a time $t=\left| 1/u_{0}-1/u_{1}\right| L/2$ with respect vertically polarized photons, where $u_{0}$ and $u_{1}$ denote the velocities of the ordinary wave and extraordinary wave in the BBO crystal, respectively. When the photon pairs exit the BBO crystal, the walk-off effect makes the state $\ket{\psi}$ becomes 
\begin{equation}
\ket{\psi'}=\frac{1}{\sqrt{2}}(\ket{0_{t}1_{0}}+e^{i\phi}\ket{1_{0}0_{t}}), \label{SPDC2}
\end{equation}
where $\ket{i_{j}}$ denotes the photon with polarized state $\ket{i}$ is delayed time $j$. The delay time $t$ makes $\ket{0_{t}1_{0}}$ and $\ket{1_{0}0_{t}}$  distinguishable. Since we use entangled polarized state in our RSP experiment, we perform a partial trace over the time system, the polarized state of photon pairs is a completely mixed state
\begin{equation}
\rho_\text{walk-off}=\frac{1}{2}(\ket{01}\!\!\bra{01}+\ket{10}\!\!\bra{10}).
\end{equation}

To compensate the photon walk-off effect, a HWP and a CBBO crystal are used in each mode $a$ and $b$. See Fig.~2 in the main text. The unitary transform of HWP for different setting angle $\theta_{\text{HWP}}$ is
\begin{equation}
U_\text{HWP}(\theta_{\text{HWP}})=  \left[ \begin{matrix}
    \text{cos}(\theta_{\text{HWP}}) & \text{sin}(\theta_{\text{HWP}}) \\
    \text{sin}(\theta_{\text{HWP}}) & -\text{cos}(\theta_{\text{HWP}})
    \end{matrix}
\right]. \nonumber
\end{equation}
By setting the angle of the HWP in each mode $a$ and $b$ at the same angle $\theta_{\text{HWP}}$, i.e.,  $\theta_{\text{HWP,a}}=\theta_{\text{HWP,b}}=\theta_{\text{HWP}}$, the output state from the BBO crystal $\ket{\psi'}$ becomes
\begin{widetext}
\begin{equation}
\begin{split}
\ket{\psi''(\theta_{\text{HWP}})}=&\frac{1}{\sqrt{2}}[(\text{cos}(\theta_{\text{HWP}})\ket{0_{t}}+\text{sin}(\theta_{\text{HWP}})\ket{1_{t}})(\text{sin}(\theta_{\text{HWP}})\ket{0_{0}}-\text{cos}(\theta_{\text{HWP}})\ket{1_{0}}) \\ &+e^{i\phi}(\text{sin}(\theta_{\text{HWP}})\ket{0_{0}}-\text{cos}(\theta_{\text{HWP}})\ket{1_{0}})(\text{cos}(\theta_{\text{HWP}})\ket{0_{t}}+\text{sin}(\theta_{\text{HWP}})\ket{1_{t}})].\label{afterHWP}
\end{split}
\end{equation}
\end{widetext}
While the photon pairs go through the CBBO crystals which make horizontally polarized photons be delayed by a time $t$ respect vertically polarized photons, the final state after the CBBO crystals is
\begin{widetext}
\begin{equation}
\begin{split}
\ket{\psi_\text{out}(\theta_{\text{HWP}})}=&\frac{1}{\sqrt{2}}[(\text{cos}(\theta_{\text{HWP}})\ket{0_{2t}}+\text{sin}(\theta_{\text{HWP}})\ket{1_{t}})(\text{sin}(\theta_{\text{HWP}})\ket{0_{t}}-\text{cos}(\theta_{\text{HWP}})\ket{1_{0}}) \\ &+e^{i\phi}(\text{sin}(\theta_{\text{HWP}})\ket{0_{t}}-\text{cos}(\theta_{\text{HWP}})\ket{1_{0}})(\text{cos}(\theta_{\text{HWP}})\ket{0_{2t}}+\text{sin}(\theta_{\text{HWP}})\ket{1_{t}})].\label{afterCBBO}
\end{split}
\end{equation}
\end{widetext}
Then, we use two QWPs and one HWP in mode $a$ to control the phase to become $e^{i\phi}=-1$. Thus, after the partial trace over the time system, the generated polarized state of the photon pairs is
\begin{widetext}
\begin{equation}
\begin{split}
\rho_\text{ent}(\theta_{\text{HWP}})&=\frac{1}{2}\text{sin}^2(\theta_{\text{HWP}})(\ket{10}-\ket{01})(\bra{10}-\bra{01})+\frac{1}{2}\text{cos}^2(\theta_{\text{HWP}})(\ket{01}\!\!\bra{01}+\ket{10}\!\!\bra{10}) \\
&=\text{sin}^2(\theta_{\text{HWP}})\ket{\Psi^{-}}\!\!\bra{\Psi^{-}}+\text{cos}^2(\theta_{\text{HWP}})\rho_\text{walk-off}, \label{rhoent}
\end{split}
\end{equation}
\end{widetext}
where $\ket{\Psi^{-}}=(\ket{10}-\ket{01})/\sqrt{2}$. We utilize $\rho_\text{ent}(\theta_{\text{HWP}})$ to calculate $\alpha_{\text{ideal}}$, $\beta_{\text{ideal}}$, and $\bar{F}_{s,\text{ideal}}(\mathcal{E})$ in Fig.~4(a) in the main text.

If the setting angle is $90^{\circ}$, the unitary
\begin{equation}
U_\text{HWP}(90^{\circ})=  \left[ \begin{matrix}
    0 & 1 \\
    1 & 0
    \end{matrix}
\right] \nonumber
\end{equation}
makes the horizontal and vertical polarization be exchanged. Thus, Eqs.~(\ref{afterHWP}) and (\ref{afterCBBO}) for $\theta_{\text{HWP}}=90^{\circ}$ are
\begin{equation}
\ket{\psi''(90^{\circ})}=\frac{1}{\sqrt{2}}(\ket{1_{t}0_{0}}+e^{i\phi}\ket{0_{0}1_{t}}) \nonumber
\end{equation}
and
\begin{equation}
\ket{\psi_\text{out}(90^{\circ})}=\frac{1}{\sqrt{2}}(\ket{1_{t}0_{t}}+e^{i\phi}\ket{0_{t}1_{t}}), \nonumber
\end{equation}
respectively. Since all photons of a pair have the same delay $pair$ which means the photon walk-off effect is compensated after the CBBO crystals, after the partial trace over the time system, for $e^{i\phi}=-1$, the generated state $\rho_\text{ent}(90^{\circ})$ in Eq.~(\ref{rhoent}) is the ideal creation of the state $\rho_\text{ent}(90^{\circ})=\ket{\Psi^{-}}\!\!\bra{\Psi^{-}}$. By contrast, if the setting angle $\theta_{\text{HWP}}=0$, the HWP do not exchanged the horizontal and vertical polarization and the photon walk-off remains after photon pairs pass through the CBBO crystals. Since the walk-off does not be compensated, the states $\ket{0_{t}1_{0}}$ and $\ket{1_{0}0_{t}}$ remain distinguishable and the generated state $\rho_\text{ent}(0)=\rho_\text{walk-off}$ is a completely mixed state.

Combining the noise model given in Eq.~(\ref{noise006}) and the polarization state of the created photon pairs varying with the photon walk-off effect as shown in Eq.~(\ref{rhoent}), the theoretical estimation of the created photon pairs is described by
\begin{equation}
\rho_\text{ent,noise}(\theta_{\text{HWP}})=0.94\rho_\text{ent}(\theta_{\text{HWP}})+0.06\frac{I\otimes I}{4}.
\label{final006}
\end{equation}
As shown in Fig.~4(a) in the main text and will be discussed in the following section, we utilize this theoretical model to calculate the theoretical values of $\alpha_{\text{theory}}$, $\beta_{\text{theory}}$, $\bar{F}_{s,\text{theory}}(\mathcal{E})$, quantum discord and EPR steerability.

\section{Implementation of RSP protocol and resulting theoretical prediction of the process steerability}


As shown in Fig.~4(a) in the main text, we realize the RSP protocol with respect to a target RSP, $\mathcal{E}^{(0)}_{\text{RSP}}$, of $U\!:~\!\!\!R(0)\!=\!(\ket{0}\!\!\bra{0}\!+\!\ket{1}\!\!\bra{0}\!+\!\ket{0}\!\!\bra{1}\!-\!\ket{1}\!\!\bra{1})\!/\!\sqrt{2}$. To obtain the theoretical prediction of the process steerability of the RSPs varying with the walk-off compensation ($\theta_{\text{HWP}}$), we utilize $\rho_\text{ent,noise}(\theta_{\text{HWP}})$ to achieve the RSPs and obtain the resulting process matrices. In experimental RSPs, the process matrices were measured by using the process tomography technique \cite{Nielsen00} and the optical setup shown in Fig.~2 in the main text. The main steps for implementing the RSP protocol are as follows.

First, as illustrated in Fig.~2 in the main text, we set the HWP which is after the phase controller and before the polarization analyzer in mode $a$ at $22.5^{\circ}$ to realize the operation $U^{\dag}$ required on Alice's side.

Second, we prepare the states $\ket{n}_{mm}\!\!\bra{n}$ as input states by the polarization analyzer consisting of a QWP, a HWP, and a polarizer in mode $a$. Note that $\ket{n}_{m}$ are eigenvectors of $\sigma_{m}$ for $n=0,1$ and $m=1,2,3$ shown in the main text. The preparation basis $\sigma_{m}$ is determined by adjusting the HWP and the QWP at different angles. For example, by adjusting the HWP and the QWP at angles of $\left( 22.5^{\circ}, 0^{\circ} \right)$ with respect to the vertical axis, the preparation basis for the polarization states is set for $\sigma_{1}$. Similarly, the preparation bases are set for $\sigma_{2}$ and $\sigma_{3}$ by adjusting the HWP and the QWP at angles of $\left( 0^{\circ}, 45^{\circ} \right)$ and $\left( 0^{\circ}, 0^{\circ} \right)$, respectively. Then, we set the polarizer at $0^{\circ}$ and $90^{\circ}$ to prepare $\ket{0}_{mm}\!\!\bra{0}$ and $\ket{1}_{mm}\!\!\bra{1}$, respectively.

Third, we perform state tomography to obtain the density matrix of output state $\mathcal{E}(\ket{n}_{mm}\!\!\bra{n})$ by the polarization analyzer in mode $b$. As mentioned in the previous step, the measurement basis is also determined by the components of the polarization analyzer in mode $b$. Taking $\ket{\bold{s}_{0}}=\ket{0}$, $\ket{\bold{s}_{0}^{\bot}}=\ket{1}$ for example, our goal is to obtain the density matrix of $\mathcal{E}(\ket{0}_{33}\!\bra{0})$. We prepare the input state $\ket{0}_{33}\!\bra{0}$ and $\ket{1}_{33}\!\bra{1}$ by adjusting the HWP, the QWP, and the polarizer at angles of $\left( 0^{\circ}, 0^{\circ}, 0^{\circ} \right)$ and $\left( 0^{\circ}, 0^{\circ}, 90^{\circ} \right)$ in mode $a$, respectively. Then, we perform the measurement of $\sigma_{1}$, $\sigma_{2}$, and $\sigma_{3}$ on $\mathcal{E}(\ket{0}_{33}\!\bra{0})$ in mode $b$ and record the coincident counts of them conditioned on $\ket{0}_{33}\!\bra{0}$ and $\ket{1}_{33}\!\bra{1}$ prepared in mode $a$. The density matrix of $\mathcal{E} \left(\ket{\bold{s}_{0}}\bra{\bold{s}_{0}}\right)$ and $\mathcal{E} \left(\ket{\bold{s}_{0}^{\bot}}\bra{\bold{s}_{0}^{\bot}}\right)$ can then be constructed through coincidence measurement results via state tomography. According to the RSP protocol, the density matrix of $\mathcal{E}(\ket{0}_{33}\!\bra{0})$ can be obtained by the mean of $\mathcal{E} \left(\ket{\bold{s}_{0}}\bra{\bold{s}_{0}}\right))$ and $\mathcal{E} \left(\ket{\bold{s}_{0}^{\bot}}\bra{\bold{s}_{0}^{\bot}}\right)$ after correction, i.e.,
\begin{equation}
\mathcal{E}(\ket{0}_{33}\!\bra{0})=\frac{1}{2}\mathcal{E} \left(\ket{\bold{s}_{0}}\bra{\bold{s}_{0}}\right))+\frac{1}{2}Z\mathcal{E} \left(\ket{\bold{s}_{0}^{\bot}}\bra{\bold{s}_{0}^{\bot}}\right)Z^{\dagger}. \nonumber
\end{equation}


Fourth, considering the theoretical measurement results of $\rho_\text{ent,noise}(\theta_{\text{HWP}})$, we can get the theoretical output states $\mathcal{E}(\ket{n}_{mm}\!\!\bra{n})$ for $n=0,1$ and $m=1,2,3$. Therefore, the theoretical process matrix $\chi_{\text{theory}}$ of the RSP can be obtained through process tomography \cite{Nielsen00}. According to $\chi_{\text{theory}}$, the process fidelity $F_{\mathcal{E},\text{theory}}$ and the process steerability $\alpha_{\text{theory}}$, $\beta_{\text{theory}}$ can be calculated via SDP \cite{Lofberg,sdpsolver}. In addition, the theoretical average state fidelity $\bar{F}_{s,\text{theory}}$ can be obtained by $\bar{F}_{s,\text{theory}}(\mathcal{E})=(2F_{\mathcal{E},\text{theory}}+1)/3$.

As shown in Fig.~4(a) in the main text, the experimental results are consistent with the theoretical predictions of $\alpha_{\text{theory}}$, $\beta_{\text{theory}}$, and $\bar{F}_{s,\text{theory}}(\mathcal{E})$. Therefore, we can conclude that $\rho_\text{ent,noise}(\theta_{\text{HWP}})$ is a good estimation of the state of the experimental photon pairs varying with the photon walk-off effect.

\section{quantum discord and EPR steerability of the created photon pairs}

In this section, we utilize the geometric discord $\mathcal{D}$ \cite{Dakic10} and steerable weight $SW$ \cite{skrzypczyk2014} to measure quantum discord and EPR steerability of the created photon pairs, respectively. We will show how to calculate the geometric discord and steerable weight in the following subsections.

\subsection{Geometric discord}

For a given $\rho_{\text{expt}}$, the geometric discord is calculated by the Bloch vector $\vec{x}\equiv\left( x_{1},x_{2},x_{3}  \right) $ with $x_{i}=\text{Tr}\left[ \rho_{\text{expt}}\left(\sigma_{i}\otimes I \right)\right]$ and the correlation tensor $A$ with $A_{ij}=\text{Tr}\left[\rho_{\text{expt}}\left(\sigma_{i}\otimes \sigma_{j}\right)\right]$ \cite{Dakic10}. Here, $\{ \sigma_{i(j)} \}^{3}_{i(j)=1}$ is composed of $\sigma_{1}=X$, $\sigma_{2}=Y$, and $\sigma_{3}=Z$ Pauli matrices. The geometric discord of $\rho_{\text{expt}}$ takes the following form
\begin{equation}
\mathcal{D}(\rho_{\text{expt}})=\frac{1}{4}\left[ \text{Tr}\left(\vec{x}^{T}\vec{x}\right)+\text{Tr}\left( A^{T}A \right)-k_{\text{max}} \right],
\label{discord}
\end{equation}
where the superscript $T$ indicates the transpose operation and $k_{\text{max}}$ is the largest eigenvalue of matrix $K=\vec{x}\vec{x}^{T}+AA^{T}$. Thus, we can obtain the experimental geometric discord, denoted as $\mathcal{D}_{\text{expt}}$, by $\rho_{\text{expt}}$.

As the results shown in Fig.~4(a) in the main text, the density matrix of the created photon pairs at $\theta_{\text{HWP}}=40^{\circ}$ in our experiment is of the following form:
\begin{widetext}
\begin{equation}
\rho_{\text{expt}}(40^{\circ})=
\left[
\begin{matrix}
0.18 & 0.13-0.04i & -0.04+0.02i & -0.01+0.09i\\
0.13+0.04i & 0.40 & -0.28+0.04i & 0.01-0.01i\\
-0.04-0.02i & -0.28-0.04i & 0.30 & -0.08+0.05i\\
-0.01-0.09i & 0.01+0.01i & -0.08-0.05i & 0.12\\
\end{matrix}
\right],
\label{rho_40}
\end{equation}
\end{widetext}
where the fidelity of $\rho_{\text{expt}}(40^{\circ})$ and the target state, $\ket{\Psi^{-}}\!\bra{\Psi^{-}}$, is $F \sim 0.63$ in our experiment. The resulting geometric discord of $\rho_{\text{expt}}(40^{\circ})$ is $\mathcal{D}_{\text{expt}}\sim 0.08$. Similarly, the theoretical geometric discord, $\mathcal{D}_{\text{theory}}\sim 0.07$, can be obtained by $\rho_\text{ent,noise}(40^{\circ})$. With $\mathcal{D}_{\text{theory}}\sim 0.07$ and $\mathcal{D}_{\text{expt}}\sim 0.08$, the state of the created photon pairs has quantum discord in terms of geometric discord, however the resulting RSP cannot satisfy the fidelity criteria (6) in the main text.

As shown in Fig.~4(b) in the main text, the density matrix of the created photon pairs at $p_{\text{noise}}=0.5$ in our experiment can be represented as
\begin{widetext}
\begin{equation}
\rho_{W,\text{expt}}(0.5)=
\left[
\begin{matrix}
0.13 & -0.01+0.01i & -0.01i & 0.01i\\
-0.01-0.01i & 0.38 & -0.23+0.01i & 0.01i\\
0.01i & -0.23-0.01i & 0.36 & 0.01-0.01i\\
-0.01i & -0.01i & 0.01+0.01i & 0.13\\
\end{matrix}
\right],
\label{rho_40}
\end{equation}
\end{widetext}
where the fidelity of $\rho_{W,\text{expt}}(0.5)$ and the target state, $\ket{\Psi^{-}}\!\bra{\Psi^{-}}$, is $F \sim  0.59$ in our experiment. The resulting geometric discord of $\rho_{W,\text{expt}}(0.5)$ is $\mathcal{D}_{\text{expt}}\sim 0.10$ and the theoretical geometric discord of $\rho_{W,\text{noise}}(0.5)$ is $\mathcal{D}_{\text{theory}}\sim 0.11$. It can be seen that the created photon pairs possess quantum discord in terms of geometric discord but the process steerability disappears. That is, it is possible that a RSP which is based on the resource of quantum discord can be described by classical RSP, $\mathcal{E}_{c}$.

\subsection{Steerable weight}
Given an assemblage $\{\rho'_{nm}\}_{nm}$ of unnormalized states which Alice's measurement affects Bob's particle into, steerable weight is defined as the minimum amount of steerable resource in $\{\rho'_{nm}\}_{nm}$ that cannot be represented in the form Eq.~(\ref{rp}).
Any assemblage $\{\rho'_{nm}\}$ can be decomposed as
\begin{equation}
\rho'_{nm}=\mu\rho'_{nm,\text{US}}+(1-\mu)\rho'_{nm,\text{S}}  \ \ \ \ \forall n,m ~,
\end{equation}
where $0\leq\mu\leq1$ and $\{\rho'_{nm,\text{US}}\}_{nm}$ ($\{\rho'_{nm,\text{S}}\}_{nm}$) is the unsteerable (steerable) assemblage in which the elements can (cannot) be represented in the form Eq.~(\ref{rp}).
The steerable weight $SW(\rho_{\text{expt}})$ is then defined as $SW(\rho_{\text{expt}})=1-\mu^*$, where $\mu^*$ denotes the maximum $\mu$ and can be obtained via SDP; see~\cite{skrzypczyk2014} for more detail. Here, to compare with quantum process steering that is measured by process matrix of the RSP protocol constructed via process tomography algorithm, we choose the measurements of the three Pauli matrices, $X$, $Y$, and $Z$ for $m=1,2,3$, respectively. Through the density matrices $\rho_{\text{expt}}$ and $\rho_\text{ent,noise}$, we can get the experimental and theoretical steerable weight, $SW_{\text{expt}}$ and $SW_{\text{theory}}$, respectively. As illustrated in Fig.~4(a) in the main text, the steerable weight of $\rho_{\text{expt}}(40^{\circ})$ shown in Eq.~({\ref{rho_40}}) is $SW_{\text{expt}}\sim 0.02$ which implies that the state of the created photon pairs has EPR steerability in terms of steerable weight, however the resulting RSP cannot satisfy the fidelity criteria (6) in the main text.

\end{document}